\documentclass[aps,superscriptaddress,twocolumn,showpacs,preprintnumbers,amsmath,amssymb,prl]{revtex4}

\usepackage{graphicx}

\newcommand{\nsum}[1]{\langle\,{#1}\,\rangle}
\newcommand{\ma}[1]{\mathcal{#1}}

\begin{document}

\title{Noise-induced macroscopic bifurcations in
 globally-coupled chaotic units}

\author{Silvia De Monte}
\affiliation{Chaos Group, The~Technical~University~of~Denmark, DK 2800 Lyngby, Denmark}
\affiliation{CEA -- Service de Physique de l'\'Etat Condens\'e,~CEN~Saclay,~91191~Gif-sur-Yvette,~France}
\author{Francesco d'Ovidio}
\affiliation{IMEDEA, CSIC University of Balearic Islands, E~07071~Palma de Mallorca, Spain}
\author{Hugues Chat\'e}
\affiliation{CEA -- Service de Physique de l'\'Etat Condens\'e,~CEN~Saclay,~91191~Gif-sur-Yvette,~France}
\author{Erik Mosekilde} 
\affiliation{Chaos Group, The~Technical~University~of~Denmark, DK 2800 Lyngby, Denmark}

\date{\today}

\begin{abstract}
Large populations 
of globally-coupled identical maps  subjected to independent additive 
noise are shown to undergo qualitative changes 
as the features of the stochastic process are varied. 
We show that for strong coupling, the collective dynamics can be described 
in terms of a few effective macroscopic degrees of freedom, whose 
deterministic equations of motion are systematically derived 
through an order parameter expansion.
\end{abstract}

\maketitle

Dynamical units coupled all-to-all constitute
relevant models of a wide range of physical and biological systems 
such as Josephson junction arrays \cite{josephson},
electrochemical oscillators \cite{kiss02}, interacting neurons
\cite{sompolinsky91} and yeast cells in a continuous-flow,
stirred tank reactor \cite{dano99}. 
Globally-coupled dynamical systems
are also widely used as prototypical models for addressing the collective
behavior of systems with many degrees of freedom \cite{mapspop}.
In the limit of strong coupling of identical units, full synchronization
usually occurs, i.e. the collective dynamics is an exact copy of the 
individual one.
However, in the generic case where some diversity and/or noise 
is present at the microscopic level, 
macroscopic measurements do not directly reflect
individual dynamical features. This simple observation is of large
experimental importance.
In the yeast cell experiment, for instance, single-cell
measurements are not available so far. The idea is thus to extract
information about metabolic processes at the individual level from
the observed collective oscillations in cell suspensions \cite{dano99}.   

In this Letter, we consider the influence of 
microscopic additive noise on the fully-synchronous regime of 
globally-coupled systems.
This allows us to isolate the effect of
noise from the possibly already complex collective dynamics
 that systems of globally-coupled identical
and noiseless units may themselves exhibit \cite{mapspop}.
Restricting, for simplicity, our discussion to identical scalar maps,
we show that noise plays a non-trivial role, ``unfolding'' the
noiseless collective dynamics 
while progressively suppressing the direct connection
between average and local evolution. 
We build a systematic expansion
that accounts {\it quantitatively} for the
collective dynamics observed, revealing that its essential features
can be described in terms of few effective degrees of freedom
and that its hierarchical structure is captured
in progressively greater detail as the truncation order is increased.

Our system of noisy and globally coupled identical
one-dimensional maps is defined as follows:
\begin{equation}\label{eq:maps} 
x_j\mapsto (1-K)\,f(x_j)+K\,\nsum{f(x)}+ \xi_j(t) \quad j=1,\dots N,
\end{equation}
where $x_j\in \mathbb R$ denotes the state of the j-th population
element, $f:\mathbb R\rightarrow \mathbb R$ is a smooth function
defining the dynamics of the uncoupled element, and
$\nsum{f(x)}:=\frac{1}{N}\sum_{j=1}^{N}\,f(x_j)$ is the average over 
the population. Every map is subjected to a white, but not necessarily
Gaussian, noise of zero average and variance $\sigma^2$.

Without noise and for strong enough coupling, 
the trajectories of all the population elements
asymptotically coincide (full synchronization). The mean field 
$X=\langle x\rangle$ then evolves on a one-dimensional manifold 
of the high-dimensional phase space of the full system.

\begin{figure}
\includegraphics[width=8.6cm,clip]{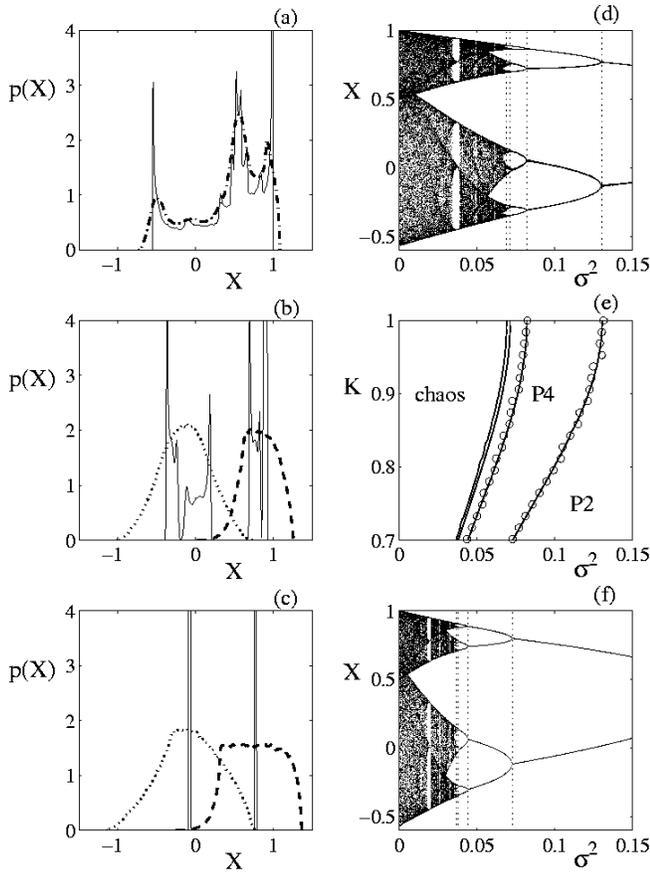}
\caption{System of $N=2^{20}$ globally coupled logistic maps 
in the chaotic regime ($f(x)=1-a\,x^2$, $a=1.57$) 
subjected to uniformly-distributed noise of variance $\sigma^2$.
(a-c): $K=0.7$: pdfs for the mean field (solid line) and one single element
(dash-dotted line; dotted line at even times, dashed line at odd times)
for different noise intensities: (a) $\sigma^2=0.0025$,  (b)
$\sigma^2=0.0324$, and (c) $\sigma^2=0.09$. 
(d): Bifurcation diagram for $X$
as a function of $\sigma^2$ for $K=1$.
(e): First period-doubling bifurcations 
for the mean field ($\circ$) 
and for the second-order reduced system (solid line, see text)
in the $(\sigma,K)$ plane.
(f): Same as (d) but for the reduced system
Eq.(\ref{eq:log2}) with $K=0.7$
(the diagram of the full system almost coincides).
The dotted lines in (d) and (f) indicate the location of the 
bifurcations reported in (e).
\label{fig:fig1}}
\end{figure}

Although noise smears any individual trajectory, for large population
sizes these fluctuations average out and the mean field $X$ 
evolves deterministically up to finite-size effects.
Somewhat counter-intuitively, the
macroscopic dynamics is qualitatively different from that of the
uncoupled map and remains apparently low-dimensional for any noise intensity.
For chaotic logistic maps, increasing noise drives the
mean field $X$ through a cascade of period-doubling
bifurcations which takes place at smaller and smaller $\sigma$ values
as the coupling strength $K$ is lowered (Fig.~\ref{fig:fig1} (d-f)).
The difference between single element noisy evolution and regular
average behaviour can
be easly seen comparing the sharply peaked probability
distribution functions (pdf) of $X$ with the broad, smooth pdfs of individual
elements (Fig. \ref{fig:fig1} (a-c)). 

\begin{figure}[h]
\includegraphics[height=.245\textwidth,clip]{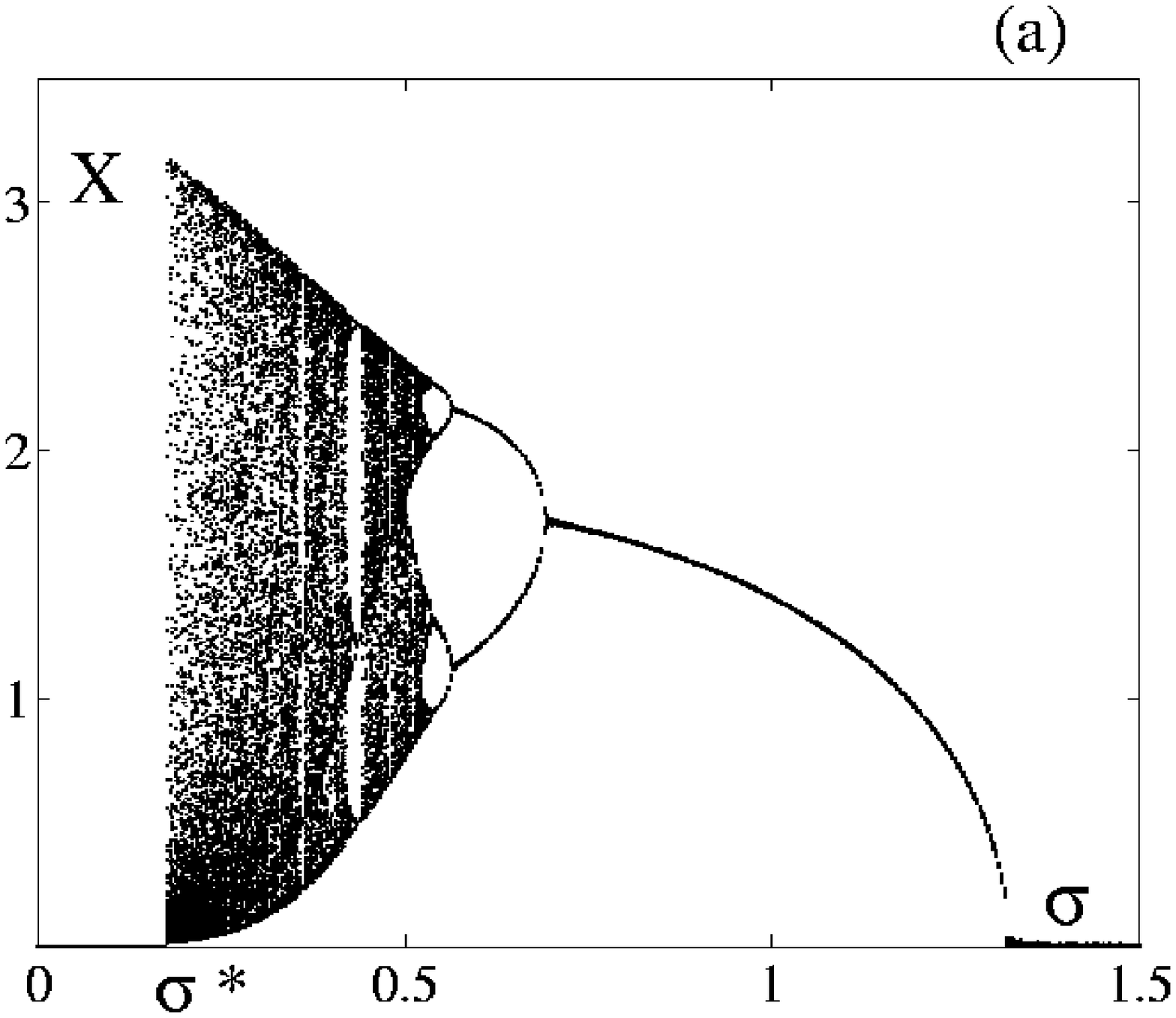}
\includegraphics[height=.245\textwidth,clip]{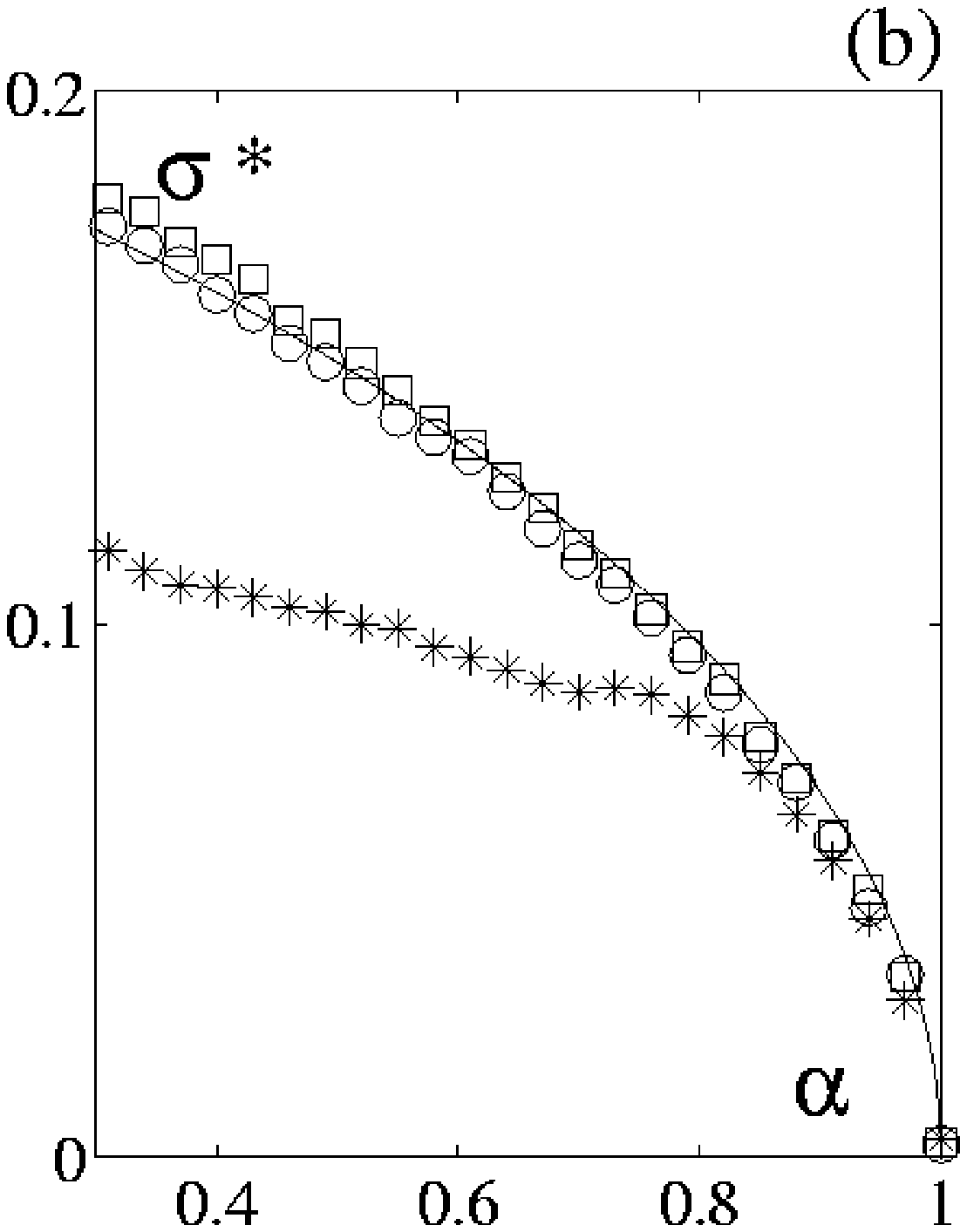}
\caption{Population of  $N=2^{22}$ excitable maps (Eq.\ (\ref{eq:exc}) with
$\alpha=0.4$, $\beta=1$, $\gamma=8$, and $K=0.9$) subjected to Gaussian
noise of variance $\sigma^2$.
(a): Bifurcation diagram for $X$ vs $\sigma$. 
(b): Critical value $\sigma^*$ of the noise intensity over
  which the mean field displays large amplitude oscillations for the
  population ($K=1$,$\square$; $K=0.8$, $\circ$; $K=0.6$,  $*$) and for
  the order parameter reduction (solid line).
 \label{fig:exc}}
\end{figure}

The effect of noise is not always the ``simplification'' of the 
collective dynamics; rather, noise unravels the underlying
complexity of the phase space.
This is particularly striking in the case of excitable
units, such as:  
\begin{equation}\label{eq:exc}f(x)=(\alpha\,x+\gamma\,x^3)\:e^{-\beta\,x^2}.
\end{equation} 
(For other examples of excitable maps, see \cite{excmaps}.)
While the origin is the unique global attractor, map (\ref{eq:exc})
possesses an unstable fixed point, excitation across which
triggers a chaotic transient.
Increasing the noise intensity over a threshold $\sigma^*$, the mean field
displays intermittent bursts, due to the crisis of a chaotic attractor which
disappears, for stronger noise, via a period-doubling cascade (Fig.\
\ref{fig:exc}).  
The onset of collective oscillations is nothing but an
explicit mechanism for ``noise-induced'' and 
``coherence'' resonance \cite{cohres}: the single-element power spectrum 
displays an enhanced peak in the periodic
regimes occurring at intermediate $\sigma$ values\cite{tbp}.

Let us now investigate analytically the apparent low-dimensional nature of the
collective motion by trying to decouple the 
macroscopic effect of noise from the dynamical ``skeleton'' provided by
the microscopic map.
We write the position of each population
element in terms of the mean field and of its displacement from it:
$x_j=X+\epsilon_j$.
The uncoupled element
equation can now be expanded as a power series around the mean field:
\begin{equation}\label{eq:exp}
f(x_j)=f(X)+\sum_{q=1}^\infty\:\frac{1}{q!}\:\ma{D}^qf(X)\:\epsilon_j^q,
\end{equation}
where $\ma{D}^qf$ indicates the $q$-th derivative of the function $f$.
Substituting Eq.\ (\ref{eq:exp}) into Eq.\ (\ref{eq:maps}), the
macroscopic first iterate is obtained by averaging:
\begin{equation}\label{eq:X}
X\mapsto \nsum{f(x)}=
f(X)+\sum_{q=1}^\infty\:\frac{1}{q!}\:\ma{D}^qf(X)\:\nsum{\epsilon^q} \;.
\end{equation}
The mean field is then coupled to a full set of new order parameters:
\begin{equation}\label{eq:ordpar}
\Omega_q:=\nsum{\epsilon^q}\hspace{20mm}q\in\mathbb N.
\end{equation}
It is worth stressing that, at this point, we have taken the infinite-size
limit, discarding the term $\nsum{\xi}$ which scales 
like $1/\sqrt{N}$
and plays the role of a macroscopic noise acting on the mean-field.
This accounts for the mean field fluctuations obeying the law 
of large numbers.\\
The evolution of the order parameters $\Omega_q$ can be computed using:
\begin{equation*}
\epsilon_j\mapsto(1-K)\:\sum_{p=1}^\infty
\:\frac{1}{p!}\:\ma{D}^pf(X)\:\left(\epsilon_j^p 
-\Omega_p\right)+\xi_j
\end{equation*}
and the fact that, in the limit $N\to \infty$, 
the positions and the noise are uncorrelated variables and, hence, 
$\nsum{h(X,\epsilon)\:\xi^q}=\nsum{h(X,\epsilon)}\:\nsum{\xi^q}$.
Eq.\ (\ref{eq:ordpar}) then yields:
\begin{eqnarray}\label{eq:omega}
\Omega_q\mapsto m_q+ &&
\sum_{i=1}^q \binom{q}{i} (1-K)^im_{q-i}\nonumber\\
&&\left\langle\left[\sum_{p=1}^\infty\frac{1}{p!}\ma{D}^pf(X)
\left(\epsilon^p-\Omega_p\right)\right]^i\right\rangle,
\end{eqnarray}
where $m_q=\nsum{\xi^q}$ is the $q$-th moment of the noise distribution.
The macroscopic dynamical system defined by Eqs.\ (\ref{eq:X}) 
and\ (\ref{eq:omega}) is still infinite-dimensional, like the full 
system, but it can be truncated to finite-order when assuming that $1-K$ 
is sufficiently small. This leads to a hierarchy of 
finite-dimensional reduced systems which capture, as we show below, 
the properties of the collective dynamics in increasing detail.

A first approximation of Eq.\ (\ref{eq:omega}) consists in keeping
only terms of zeroth order in $(1-K)$, so that any $\Omega_q=m_q$. 
The scalar equation for the mean field:
\begin{equation}\label{eq:X0}
X\mapsto 
f(X)+\sum_{q=1}^\infty\:\frac{1}{q!}\:\ma{D}^qf(X)\:m_q
\end{equation} 
describes exactly the bifurcation diagram for $K=1$ Fig.\ \ref{fig:fig1} (d).
Here, the macroscopic effect of noise is a 
one-dimensional unfolding of the uncoupled element dynamics, 
whose nonlinearities become progressively more relevant 
as the noise intensity increases.   
In particular, if the map is polynomial, only a finite number of moments
influence the mean-field dynamics. This suggests a criterium for 
dividing the noise distributions into classes according to the collective
behaviour they generate. 

\begin{figure}
\includegraphics[height=.5\textwidth,clip,angle=-90]{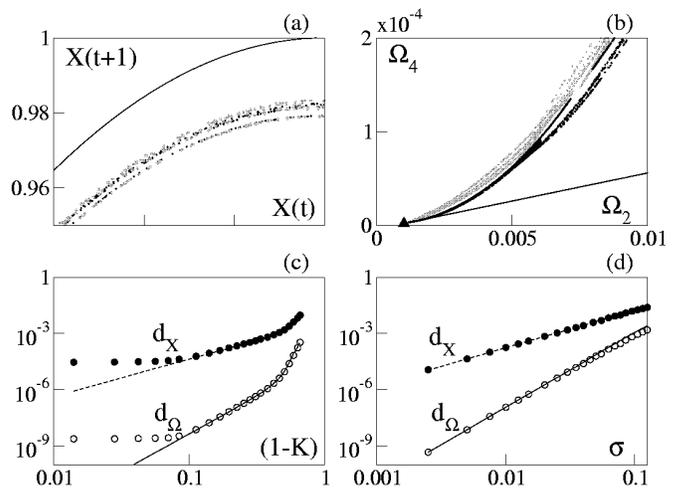}
\caption{Population of Fig. \ref{fig:fig1}
(a): First return map of $X$ for the full system (dark dots),
its zeroth-order approximation (solid line), and its second-order
approximation (light dots) ($K=0.4$ and $\sigma^2=0.001$).
(b):  $\Omega_4$ versus $\Omega_2$ for the same parameters (dark dots: full system,
zeroth-order: triangle, second-order: solid line, fourth-order: light dots).
(c) and (d) distances $d_x$ and $d_\Omega$ defined in text
(circles for the full system, dashed lines for the second-order approximation,
solid line for fourth-order system).
((c): $\sigma^2=0.03$; (d): $K=0.6$).\label{fig:fig2}}
\end{figure}

For logistic maps, Eq.\ (\ref{eq:X0}) takes the simple form:
\begin{equation}\label{eq:log0}
X\mapsto 1-a\:\sigma^2-a\:X^2.
\end{equation}
This map is easily rescaled to a logistic map
in which $\sigma^2$ plays the role of the usual nonlinearity parameter.
Although Eq.\ (\ref{eq:log0}) provides a good approximation to the regimes of
very strong coupling, it is independent of $K$. Hence, it cannot account 
for the changes with $K$ of the period-doubling
bifurcation lines (Fig.~\ref{fig:fig1}(e)). 
Truncating  Eqs.\ (\ref{eq:X}) and\ (\ref{eq:omega}) to the second order
in $(1-K)$ (the first order term vanishes for zero-average noise)
the coupling strength appears as a parameter for the macroscopic equations
of motion. In this case,
the higher-order moments obey a recurrence relation leading, when $f$ is
polynomial, to a two-dimensional map. For our example of logistic maps, this
reads:  
\begin{eqnarray}\label{eq:log2}
X\mapsto && 1-a\,X^2-a\:\Omega_2\nonumber\\
\Omega_2\mapsto && \sigma^2+(1-K)^2a^2
\\&& \times
\left[m_4-6\sigma^4+(4X^2-\Omega_2+6 \sigma^2)\Omega_2\right].
\nonumber
\end{eqnarray}    
Figure~\ref{fig:fig1}(e) shows that even for intermediate coupling 
strengths this equation accounts quantitatively for the main features 
of the mean field dynamics. \\
Looking at the bifurcation cascade of Fig.\
\ref{fig:fig1} (f), one could think that Eqs.\ (\ref{eq:log2}) might be
further reduced to a scalar map. This is not possible, however, as the mean
field first return map is folded (Fig.\ \ref{fig:fig2}) and hence
the macroscopic system must have more than one variable.
Remarkably, our approximation
at order $(1-K)^2$ (Eq.~(\ref{eq:log2}) for logistic maps) 
reproduces this folded structure.
For both the full system and the second-order approximation
Eq.~(\ref{eq:log2}), the distance $d_X$  between the first return map of $X$
and the  parabola identified by the zeroth order truncation scales, as
expected, like $(1-K)^2$ or $\sigma^2$  (Fig.\ \ref{fig:fig2}(c,d)) and
vanishes on $K=1$, where the zeroth-order truncation Eq.~(\ref{eq:log0}) is
exact.  
The ``thickness'' of the transverse folded structure
shown in Fig.~\ref{fig:fig2}(a) scales similarly \cite{tbp},
indicating that, apart from finite-size effects that may prevent us
from observing it, the dimension of the macroscopic attractor becomes
larger than that of the uncoupled element as soon as $K<1$ and $\sigma>0$.
\\
Going one step further, it is easily seen that there are still
features of the mean field dynamics that cannot be 
captured by Eq.\ (\ref{eq:log2}). The macroscopic variables 
$\Omega_q$ for the population satisfy only approximately the relations 
used for reducing the dimensionality of the truncated system. 
For instance, $\Omega_4$ displays a complex dependence on $\Omega_2$
(Fig.~\ref{fig:fig2}(b)), while it is constant at zeroth order,
and linear at second-order. In turn, this complex structure
is retrieved if Eqs.~(\ref{eq:X}) and (\ref{eq:omega}) 
are truncated to fourth order in $(1-K)$.  
In the $(\Omega_2,\Omega_4)$ plane, 
the distances $d_\Omega$ separating the second-order truncation 
from the full system or the fourth-order truncation are comparable
and they scale like the expected $(1-K)^4$ and $\sigma^4$
(Figs.~\ref{fig:fig2}(c,d)).

Extrapolating from the above results, we conclude that the presence of
noise unfolds the fully-synchronized
regime into an infinitely complex but hierarchically organized
structure whose main features are captured by the lowest orders
of our approximation scheme.
Nevertheless, in the parameter region considered here, we expect the 
collective chaos to remain embedded in a finite dimensional space.
This is perhaps best seen when calculating the Lyapunov
exponents of the finite-dimensional maps approximating the full system.
The maps possess only one positive exponent in the chaotic region, 
whereas increasing the order of the
approximation adds ``new'', more negative exponents to those existing
at previous order. These exponent values agree well with those 
numerically computed by simulating the Perron-Frobenius dynamics of the
population pdf \cite{tbp}, so that we are confident that they 
do represent the collective dynamics of the full system. 

In general, when the local map $f$ is not polynomial,
the order parameter expansion of Eq.\ (\ref{eq:exc}) contains an
infinite number of terms even for $K=1$, and it is then necessary
to introduce the further closure assumption $\sigma\ll 1$ to obtain 
closed form approximations. We can illustrate this for the excitable map
example mentioned above. At second order in both $(1-K)$ and $\sigma$,
the approximation accounts well for the establishment of 
collective oscillations. 
The linear stability analysis of the origin for the reduced system 
provides an analytical condition for the onset of the intermittent 
behavior, which is in good agreement with the simulations of the population 
(Fig.\ \ref{fig:exc}{ (b)}).

To summarize, the collective dynamics of infinitely-large populations 
of globally-coupled identical maps subjected to independent additive 
noise is deterministic and undergoes qualitative changes 
as the features of the stochastic process are varied. 
For strong coupling, the collective dynamics can be described 
in terms of a few effective macroscopic degrees of freedom, whose 
equations of motion 
have been systematically derived through an order parameter expansion 
valid for any smooth map and noise distribution. 
Such a reduced system allows us to understand how the nonlinearities of the 
map interact with the specific features of the microscopic noise.
At the experimental level, our work indicates that the observation of
unusual pdfs of local variables
such a those shown in Fig.~\ref{fig:fig1} may be traced back to the
presence of an underlying chaotic local dynamics.

Our approach reveals that in a large region of parameter space
the collective motion, in spite of being endowed with a single
positive Lyapunov exponent, takes place on a hierarchically-organized complex
attractor. 
We believe that this complexity is but another facet of the ``anomalous
scaling'' properties of the cloud of points representing the population
in the local phase space which were uncovered recently by Teramae
and Kuramoto \cite{teramae01}, both phenomena being rooted in the
structure of the global phase space around the fully-synchronized solution.

The method proposed here, together with Ref.\ \cite{our}, allows 
to compare the macroscopic bifurcations induced by 
different sources of microscopic disorder, such as noise and parameter
mismatch (quenched noise). 
In this last case, an order parameter expansion with other closure
assumptions provides a different low-dimensional unfolding of the 
local dynamics, accounting for the observed diversity of the bifurcation
scenarios \cite{tbp}. 

Future work should consider how all these aspects evolve as one lowers the
coupling 
strength towards the weak-coupling regimes considered by Shibata et al.
\cite{shibata99} where collective chaos is high-dimensional. Our work can be
extended to the case of continuous-time systems, where we expect
a similar effect of microscopic noise \cite{continuous}.
Finally, it would be most
interesting to make the connection with  lattices or networks of
locally-coupled maps, and the  so-called non-trivial collective behavior they
exhibit. There, as suggested in \cite{lemaitre}, the equivalent of the noise
considered here lies in the finite-amplitude fluctuations introduced by the
neighbors and the finite correlation length in the system.

S.D.M. acknowledges
support from the ESF Programme REACTOR. The authors are grateful to
the referees for their comments.

\end{document}